# Cross-layer Resource Allocation Scheme for Multi-band High Rate UWB Systems

Ayman Khalil, Matthieu Crussière and Jean-François Hélard

*Institute of Electronics and Telecommunications of Rennes (IETR)*
*INSA, 20 avenue des Buttes de Coësmes, 35043 Rennes, France*
ayman.khalil@insa-rennes.fr

*Abstract-* **In this paper, we investigate the use of a cross-layer allocation mechanism for the high-rate ultra-wideband (UWB) systems. The aim of this paper is twofold. First, through the cross-layer approach that provides a new service differentiation approach to the fully distributed UWB systems, we support traffic with quality of service (QoS) guarantee in a multi-user context. Second, we exploit the effective SINR method that represents the characteristics of multiple sub-carrier SINRs in the multi-band WiMedia solution proposed for UWB systems, in order to provide the channel state information needed for the multi-user sub-band allocation. This new approach improves the system performance and optimizes the spectrum utilization with a low cost data exchange between the different users while guaranteeing the required QoS. In addition, this new approach solves the problem of the cohabitation of more than three users in the same WiMedia channel.**

## I. Introduction

Ultra-wideband (UWB) transmission has attracted significant interest since 2002 when the Federal Communications Commission (FCC) regulated UWB systems by allocating the 3.1 to 10.6 GHz spectrum for unlicensed use of UWB [1]. In order to reduce interference with other existing systems, the FCC imposed a power spectral density (PSD) limit of -41.3 dBm/MHz.

The IEEE 802.15.3a wireless personal area networks (WPAN) standardization group defined a very high data rate physical layer based on UWB signalling. One of the multiple-access techniques considered by the group is a multi-band orthogonal frequency division multiplexing (MB-OFDM) supported by the MultiBand OFDM Alliance (MBOA) and the WiMedia forum [2], [3], which merged in March 2005 and are today known as the WiMedia Alliance.

On December 2005, ECMA International eventually approved two standards for UWB technology based on the WiMedia solution: ECMA-368 for High Rate Ultra Wideband PHY and MAC Standard and ECMA-369 for MAC-PHY Interface for ECMA-368.

To this date, most research studies on multi-band UWB systems have been devoted to the physical layer issues. In [4]–[6], the authors propose different sub-band and power allocation strategies based on the channel information without taking into account the users requirements and the quality of service issues. On the other hand, in [7] and [8], the authors propose scheduling and power allocation algorithms that provide quality of service for multimedia applications in UWB systems but without having a full knowledge of the channel.

The aim of this paper is to propose a cross-layer resource allocation mechanism based on UWB signalling in a multi-user context. The proposed scheme exploits scheduling and sub-band allocation principles to maintain an efficient use of the spectrum in a multiple medium access demand. In our work, we consider the WiMedia solution for the PHY and MAC layers of the UWB systems. In the PHY layer, we exploit the effective SINR method proposed recently in the 3GPP standardization which can be effectively used in multi-band OFDM systems in order to compute an adequate metric to be forwarded to the MAC layer. It consists in representing the channel state information through a single value that is strongly correlated with the actual BER. In the MAC layer, we define a service differentiation scheme and a scheduler in order to control each user requirements, to allow the cohabitation of more than three users in the same channel and to differentiate between two major traffic classes: hard-QoS and soft-QoS. The definition of these two classes is useful for UWB systems, where multimedia or real-time applications (video recording, A/V conferencing, etc) should have a certain priority on data or non real-time applications (file transfer, wireless USB, etc).

The remainder of this paper is organized as follows. Section II introduces the system model by presenting the PHY and MAC layers conditions. Section III details the proposed scheme by introducing the new PHY and MAC layers functionalities and the new cross-layer optimization scheme. Section IV presents simulation results showing the efficiency of the new scheme, how it outperforms the WiMedia solution for hard-QoS users without decreasing significantly the performance of other users. Finally, section V concludes this paper.

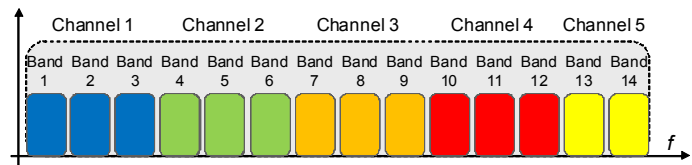

Fig. 1. Channel distribution for WiMedia solution.

## II. SYSTEM MODEL – WIMEDIA SOLUTION

### A. PHY Layer

The WiMedia solution consists in combining OFDM with a multi-banding technique that divides the available band into 14 sub-bands of 528 MHz, as illustrated in Fig. 1. An OFDM signal can be transmitted on each sub-band using a 128-point inverse fast Fourrier transform (IFFT). Out of the 128 subcarriers used, only 100 are assigned to transmit data. Different data rates from 53.3 to 480 Mbit/s are obtained through the use of forward error correction (FEC), frequency-domain spreading (FDS) and time-domain spreading (TDS), as presented in Table I. The constellation applied to the different subcarriers is either a quadrature phase-shift keying (QPSK) for the low data rates or a dual carrier modulation (DCM) for the high data rates. Time-frequency codes (TFC) are used to provide frequency hopping from a sub-band to another at the end of each OFDM symbol. TFC allows every user to benefit from frequency diversity over a bandwidth equal to the three sub-bands of one channel. In addition, to prevent from interference between consecutive symbols, a zero padding (ZP) guard interval is inserted instead of the traditional cyclic prefix (CP) used in the classical OFDM systems [9].

The WiMedia solution offers potential advantages for high-rate UWB applications, such as the signal robustness against channel selectivity and the efficient exploitation of the energy of every signal received within the prefix margin. However, we will see in the next section that the exploitation of the PHY layer at the MAC level is suboptimal in a multi-user context since the medium access mechanisms do not take advantage of the sub-band dimension which should be used to multiplex up to three applications in the frequency domain.

### B. MAC Layer

The WiMedia MAC protocol is a distributed TDMA-based MAC protocol as defined in ECMA standard [10]. Time is divided into superframes where each frame is composed of 256 medium access slots (MAS) as illustrated in Fig. 2. Each MAS has a length of 256 µs. Each superframe starts with a beacon period (BP) that is responsible for the exchange of reservation information, the establishment of neighbourhood information and many other functions.

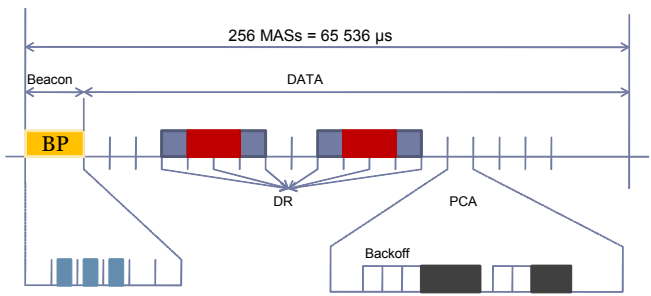

Fig. 2. MAC superframe structure.

TABLE I
WiMedia system data rates

| Data Rate (Mbit/s) | Modulation | Coding Rate | FDS | TDS |
|---|---|---|---|---|
| 53.3 | QPSK | 1/3 | Yes | Yes |
| 80 | QPSK | 1/2 | Yes | Yes |
| 110 | QPSK | 11/32 | No | Yes |
| 160 | QPSK | 1/2 | No | Yes |
| 200 | QPSK | 5/8 | No | No |
| 320 | DCM | 1/2 | No | No |
| 400 | DCM | 5/8 | No | No |
| 480 | DCM | 3/4 | No | No |

WiMedia defines two access mechanisms: the prioritized contention access (PCA) and the distributed reservation protocol (DRP).

PCA provides differentiated access to the medium for four access categories (ACs); it is similar to the enhanced distributed channel access (EDCA) mechanism of IEEE 802.11e standard [16]. On the other hand, DRP is a TDMA-based mechanism which enables a device to reserve one or more MASs for the communication with neighbours.
ECMA defines two types of reservation: hard reservation and soft reservation. In the hard reservation case, devices other than the reservation owner and target(s) shall not transmit frames; that means that unused time should be released for PCA. On the other hand, the soft reservation type permits PCA, but the reservation owner has preferential access.

In brief, any of the defined mechanisms is based on an efficient service differentiation that can guarantee a certain level of QoS for strict QoS applications. The main disadvantage of PCA mechanism is the collision that could happen between the users due to the use of random values to access the medium (backoff and contention window). On the other hand, DRP mechanism solves the problem of collision, but it is not based on a service differentiation principle.

In order to join the advantages of the two mechanisms, we define in our work a new mechanism based on the DRP principle regarding the negotiation and reservation, and capable of providing at the same time, a service differentiation based on a simple approach as detailed in the next section.

On the other hand, all the medium access mechanisms specified above follow a pure TDMA approach that completely excludes any exploitation of the multi-band characteristics of the MB-OFDM signaling used at the PHY level, since the TFC is not exploited to handle the multi-user aspects. This means that each transmitted signal occupies only one third of the available bandwidth at each symbol time slot, whatever the TFC used. Our proposed medium access mechanism will hence be based on a full exploitation of the resource at the PHY level, allowing a sub-band spectrum sharing between users, as well as a time slot allocation through scheduling principles.

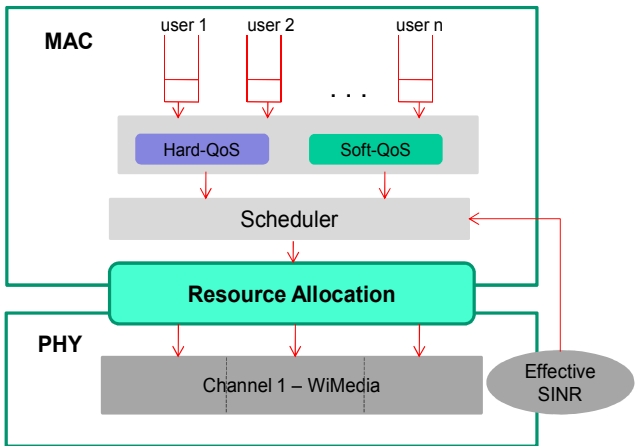

Fig. 3. Cross-layer approach.

## III. PROPOSED SCHEME

We propose in this section a new medium access approach based on a sub-band allocation and scheduling for the multi-user multi-band UWB systems. The medium access is managed by combining information provided by the PHY and MAC layers in order to form a cross-layer scheme. The concept of the proposed scheme is illustrated in Fig. 3. As depicted, the proposed scheme takes into account the users requirements and constraints as well as the users channel power of each sub-band. Hence, based on the ECMA standard that defines the multi-band UWB physical and MAC specifications, we define enhanced physical and MAC functionalities capable to work jointly in an optimized way in order not to increase the system complexity and to reduce the cost of exchanged data between devices in a distributed architecture.

### A. At the PHY layer

WiMedia solution specifies a multi-band OFDM scheme to transmit information, where each channel is divided into three sub-bands and the allocation is made by sub-band. For instance, three users in one channel have to share the available three sub-bands. Hence, to provide the users channel powers of each sub-band in order to achieve efficient spectrum utilization and a sub-band allocation that respects each user PHY conditions, the channel state information (CSI) is needed at the transmitter side. To do so, we propose to use the effective SINR method to represent the characteristics of each sub-band and to evaluate the system level performance after channel decoding in terms of BER. This can be motivated, from the physical point of view by the need of such measures for accurate and realistic evaluation of the system level performance but also for suitable development of link adaptation algorithms such as adaptive modulation and coding, packet scheduling, etc.

The effective SINR method consists in finding a compression function that maps the sequence of varying SINRs to a single value that is strongly correlated with the actual BER [11]. It is given by

$$SINR_{eff} = I^{-1}\left(\frac{1}{N}\sum_{i=1}^{N} I(SINR_i)\right) \quad (1)$$

In the effective SINR method, as in [11] and [12], we use the following information measure function $I(x)$:

$$I(x) = \exp(-\frac{x}{\lambda}) \quad (2)$$

The inverse function of $I(x)$:

$$I^{-1}(x) = -\lambda \ln(x) \quad (3)$$

Thus,

$$SINR_{eff} = -\lambda \ln\left[\frac{1}{N}\sum_{i=1}^{N} \exp(-\frac{SINR_i}{\lambda})\right] \quad (4)$$

where $\lambda$ is a scaling factor that depends on the selected modulation and coding scheme (MCS) [13], $N$ the number of subcarriers in a sub-band, and $SINR_i$ the ratio of signal to interference and noise in the $i^{th}$ subcarrier.

In our system model, we compute the effective SINR value for each user in each sub-band by using (4). For instance, in the case of one channel divided into $N_b = 3$ sub-bands, and with $N_u = 3$ users, the computation result is a matrix containing $N_b \times N_u = 9$ effective SINR values.

### B. At the MAC layer

As presented in section II, the MAC layer of WiMedia is either a distributed, reservation-based channel access mechanism (DRP) or a prioritized, contention-based channel access mechanism. In order to provide quality of service to strict QoS applications and to prevent collision between users, we define a new mechanism based on the reservation and a service differentiation at the same time. The reservation is achieved with the same manner as presented in ECMA. However, the service differentiation is achieved via a new service classifier entity whose task is to classify all the incoming traffics into two traffic types: hard-QoS class for applications that have strict requirements and soft-QoS class for applications that have tolerance for some requirements. For example, a real-time application (e.g. video streaming) has a lower tolerance for delay than non real-time application (data transfer).

The service classifier entity is based on the weight or the priority level concept. Hence, each user is assigned a weight $q$ that depends on its requirements in terms of throughput, delay and error rate. Consequently, hard-QoS users should be assigned a weight greater than the one assigned to the soft-QoS users.

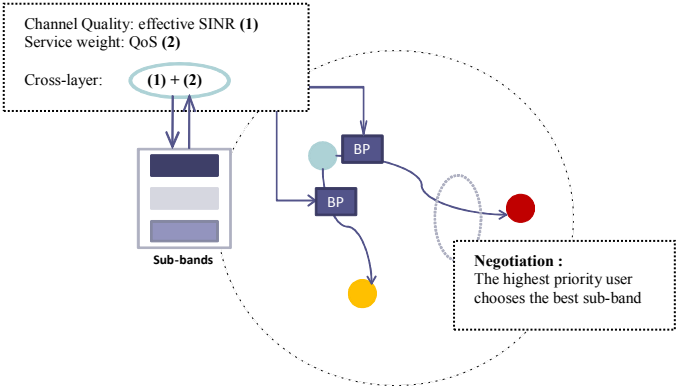

Fig. 4. Cross-layer in the distributed architecture.

The weight $q$ is assigned in a way that respects the following conditions

$$\begin{cases} \sum_{i=1}^{U} q_i = 1 \\ q_{hard-QoS} = k \times q_{soft-QoS} \end{cases} \quad (5)$$

where $U$ is the total number of users and $k$ a positive constant greater than one which value depends on the ratio of hard-QoS users number to soft-QoS users number.

### C. Cross-layer Algorithm

As mentioned before, the aim of the proposed scheme is to enhance the spectrum utilization via a cross-layer mechanism that should be able to combine the new PHY and MAC functionalities in an optimized way in order to improve the system performance without increasing its complexity. Our main objective is to improve each hard-QoS user performance by minimizing its BER, while maintaining the MAC layer conditions under the defined allocation policy. In addition, when the number of users is greater than three, the cross-layer scheme should act as a scheduler that is capable of sharing the available sub-bands between the existing users with an efficient way that respects all the requirements. To do so, we follow the principle of guaranteeing a certain amount of resource for high-priority users and sharing the remaining amount among low-priority users. For example, in the case of four users, the two lowest-priority users should share the same sub-band, which causes some performance degradation for soft-QoS users that are forced to use half of the allocated rate. Consequently, these users will have a delay which is two times greater than the hard-QoS users delay. This degradation in soft-QoS users performance is acceptable due to the fact that the non real-time applications are tolerant for the delay.

Therefore, we define the cross-layer function as an allocation level ($AL$) function that shares the available sub-bands between the existing users in a priority-based approach. For instance, the user that has the highest $AL$ is assigned the most powerful sub-band. The allocation level function of user $U_n$ is given by

$$AL(n) = W_{MAC} q_{U_n} + W_{PHY} \max_i (SINR_{eff}(U_n, B_i)) \quad (6)$$

where $q_n$ represents the weight or the priority level of user $U_n$, $SINR_{eff}(U_n, B_i)$ the effective SINR of user $U_n$ in band $B_i$. $W_{MAC}$ and $W_{PHY}$ are two constants that represent the balance between MAC and PHY layers. For instance, giving a zero value for one of these constants will cancel the function of its layer.

In order to optimize the bandwidth utilization and to reduce the amount of exchanged data between the devices in the distributed architecture, the result of the cross-layer algorithm given by (6) and the choice of the corresponding sub-bands are transmitted by the device having traffic to send to all the existing devices in its piconet via the information elements (IEs) used in the BP as given in ECMA standard (see Fig. 4). Hence, each device calculates its $AL$ and determines its sub-band sequence in a preferred order. Consequently, at each superframe, the sub-bands are allocated according to the $AL$ and the sub-band sequences computed by each user; for example, [2 3 1] is a sub-band sequence for a user having the highest power (greatest $SINR_{eff}$) in sub-band 2 and the lowest power in sub-band 1. Thus, the $AL$ and the sub-band sequence are sent and shared between users via the IEs during the BP in order to make an efficient scheduling.

Besides, the allocation is updated at the beginning of each superframe. Accordingly, this allocation strategy is useful and can be efficiently applied for indoor UWB communications without significantly increasing the system complexity, thanks to the slow time variations of the UWB channel.

The negotiation takes place whenever two or more devices choose the same sub-band in the case of three users scheme, or should share the same sub-band in the case of more than three users scheme. Thus, the most powerful user, i.e. the user that has the greatest $AL$, is assigned its highest priority sub-band, and the second most powerful user has to choose its second highest priority sub-band. Consequently, in the case of three users in one channel, the least powerful user is assigned the remaining sub-band. After the negotiation, the reservation of MASs is performed with the same manner as in the DRP mechanism.

### IV. SYSTEM PERFORMANCE

#### A. Channel Model

The channel used in this study is the one adopted by the IEEE 802.15.3a committee for the evaluation of UWB physical layer proposals [14]. This model is a modified version of Saleh-Valenzuela model for indoor channels [15], fitting the properties of UWB channels. A log-normal distribution is used for the multipath gain magnitude. In addition, independent fading is assumed for each cluster and each ray within the cluster. The impulse response of the multipath model is given by

$$h_i(t) = G_i \sum_{z=0}^{Z_i} \sum_{p=0}^{P_i} \alpha_i(z,p) \delta(t - T_i(z) - \tau_i(z,p)) \quad (7)$$

where $G_i$ is the log-normal shadowing of the $i^{th}$ channel realization, $T_i(z)$ the delay of cluster $z$, and $\alpha_i(z,p)$ and $\tau_i(z,p)$ represent the gain and the delay of multipath $p$ within cluster $z$, respectively.

TABLE II
Characteristics of UWB channels

|  | CM1 | CM2 | CM3 | CM4 |
|---|---|---|---|---|
| Mean excess delay(ns) | 5.05 | 10.38 | 14.08 | - |
| RMS delay spread (ns) | 5.28 | 8.03 | 14.28 | 25 |
| Distance (m) | <4 | <4 | 4-10 | 10 |
| LOS/NLOS | LOS | NLOS | NLOS | NLOS |

Four different channel models (CM1 to CM4) are defined for the UWB system modelling, each with arrival rates and decay factors chosen to match different usage scenarios and to fit line-of-sight (LOS) and non-line-of-sight (NLOS) cases. The channel models characteristics are presented in Table II.

### B. Simulation Results

In this section, we present the simulation results for the proposed cross-layer allocation scheme and we compare the performance of the new scheme with the performance of WiMedia solution using TFC. Therefore, we use the proposed WiMedia data rates (see Table I). The results are performed on the first three WiMedia sub-bands (3.1-4.7 GHz) for CM1 channel model.

In Fig. 5, we present the case of three users transmitting simultaneously in the first WiMedia channel. The three users are assigned different data rates in order to show the advantage of hard-QoS users on soft-QoS users in term of error rate. We have one hard-QoS user transmitting at the highest rate, i.e. 480 Mbps, and two soft-QoS users transmitting at 400 Mbps. We can see that the hard-QoS user outperforms the soft-QoS users with a considerable gain although it is transmitting at a higher rate. Hence, this gain proves that the proposed scheme can guarantee a higher performance for multimedia users even if they have strict requirements.

In Fig. 6, we compare the performance of three users scheme transmitting at a data rate of 320 Mbps with the performance of a single-user adopted by WiMedia solution transmitting at the same data rate. Note that for the single user solution, TFC is exploited for the comparison because it

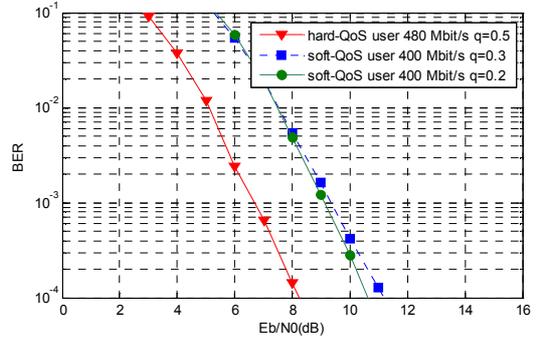
Fig. 5. Performance of three users transmiting with different data rates.

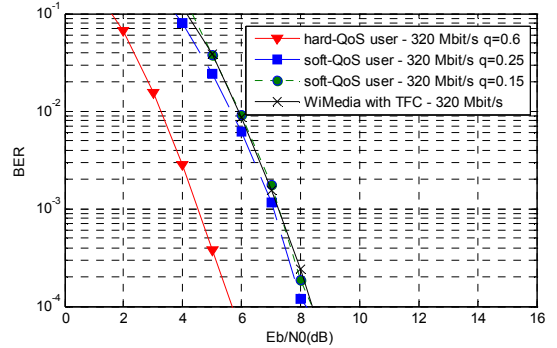
Fig. 6. Performance of the cross-layer scheme compared to WiMedia solution with TFC.

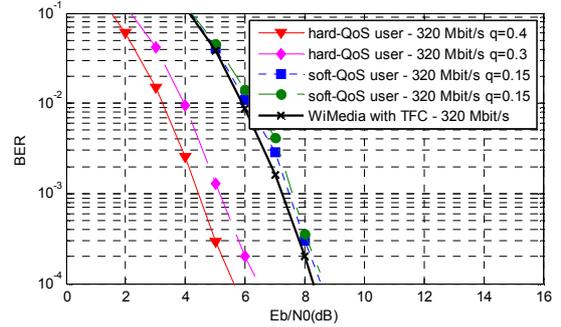
Fig. 7. Performance in a four users scheme.

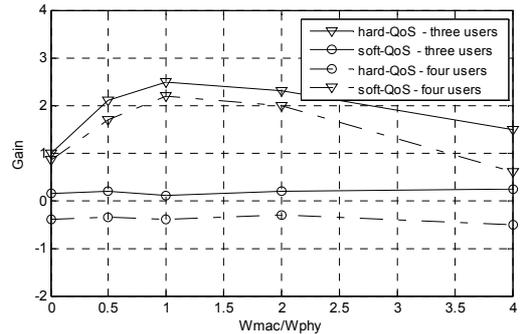
Fig. 8. MAC-PHY balance.

offers better performance. As show in the figure, for a $BER = 10^{-4}$, the cross-layer scheme offers a 2.5 dB gain for the hard-QoS user compared to WiMedia solution. The performance of soft-QoS users is close to that of WiMedia solution.

In Fig. 7, we present the performance of the four users scheme transmitting simultaneously in the first channel at the same rate of 320 Mbit/s; two hard-QoS users with two different weights and two soft-QoS users with the same weight. In the same channel, the best sub-bands are guaranteed for the two hard-QoS users respectively and the remaining sub-band is shared between the soft-QoS users. Hence, note that the performance of the soft-QoS users is slightly degraded compared to WiMedia single user performance.

In Fig. 8, we show the influence of the layers weights $W_{PHY}$ and $W_{MAC}$ defined in (6). Indeed, we show that a good balance between the two layers gives the best performance. For instance, if we cancel the MAC function, i.e. for $W_{MAC}/W_{PHY} = 0$, we note that the average gain is not that considerable. On the other hand, increasing $W_{MAC}/W_{PHY}$ will decrease the gain value. A good compromise is to balance between the two layers in order to achieve the best performance.

## V. CONCLUSION

In this paper, we proposed a new approach for the resource allocation under QoS requirements for the next generation multi-band high data rate UWB systems. This new approach is based on a cross-layer scheme which combines information provided by the PHY and MAC layers in order to achieve an efficient and optimized sub-band allocation in a distributed multi-user access, so that hard-QoS users have advantage on soft-QoS users in term of error rate. We showed that the proposed scheme respects the distributed MAC architecture, so that the amount of exchanged information required to make all the devices follow the same cross-layer policy is very low. Besides, the new scheme solves the problem of the cohabitation of more than three users in one WiMedia channel without increasing the system complexity.

ACKNOWLEDGMENT

The research leading to these results has received funding from the European Community's Seventh Framework Programme FP7/2007-2013 under grant agreement n° 213311 also referred as OMEGA.